\def\Journal#1#2#3#4{{#1} {\bf #2}, #3 (#4).}

\def\NPB{{ Nucl. Phys.} B}

\def\PRD{{ Phys. Rev.} D}

\documentclass[aps,12pt,floatfix,tightenlines,showkeys,showpacs]{revtex4}
\usepackage{graphicx}
\newcommand{\bfr}{{\bf r}}

\newcommand{\ben}{\begin{displaymath}}
\newcommand{\een}{\end{displaymath}}
\newcommand{\be}{\begin{equation}}
\newcommand{\ee}{\end{equation}}
\newcommand{\bea}{\begin{eqnarray}}
\newcommand{\eea}{\end{eqnarray}}

\newcommand{\eq}[1]{Eq.~(\ref{#1})}

\newcommand{\bfp}{{\bf p}}

\newcommand{\bfq}{{\bf q}}

    \newcommand {\boldsigma}{\mbox{\boldmath$\sigma$}}

\begin{document}

\title{\bf  \hskip10cm NT@UW-13-28\\
{ Pionic and Hidden-Color, Six-Quark  Contributions to the  Deuteron $b_1$ Structure Function}}

\author{ Gerald A. Miller }

\affiliation{Department of Physics,
University of Washington, Seattle, WA 98195-1560}

\date{\today}

\begin{abstract}

{The  $b_1$ structure function is  an observable feature   of a spin-1 system  sensitive to   non-nucleonic components of  the target nuclear wave function. The contributions of exchanged pions in the deuteron are estimated and  found to be  of measurable size for small values of $x$. A simple model for a hidden-color, six-quark configurations (with $\sim 0.15\%$  probability to exist in the deuteron)  is proposed and found to give substantial contributions 
 for values of $x>0.2$. Good  agreement with Hermes data  is obtained. Predictions are made for an upcoming JLab experiment. The   Close \& Kumano sum rule is  investigated and  found to be a useful guide to understanding   various possible effects that may contribute. }
\end{abstract}\pacs{nn??}

\maketitle     
\noindent
  \def\quup{q^1_{\uparrow}}
\def\qudp{q^{1}_{\downarrow}}
\def\quum{q^{-1}_{\uparrow}}
\def\qudm{q^{-1}_{\downarrow}}
\def\quuz{q^0_{\uparrow}}
\def\qudz{q^0_{\downarrow}}
  \section{Introduction}
  Deep inelastic scattering from a spin-one target has features, residing in the leading-twist  $b_1$ structure function,  that are not  present for a spin-1/2 target~\cite{jaffe,frankfurt}.
  In the Quark-Parton model  
  \bea b_1= \sum_q e_q^2 \left[\quuz-{1\over2}(\quup+\quum)\right]\,\equiv\sum_ie_q^2\delta q_i\label{pm1}
  \eea
where $q^{m}_{\uparrow}$($q^{m}_{\downarrow}$) is the number density of 
quarks with spin up(down) along the $z$ axis in a target hadron with
helicity $m$.  The function $b_1$ is called the tensor structure function of the deuteron because it has been observed using a tensor polarized deuteron target~\cite{Airapetian:2005cb} for  values of Bjorken $0.01<x<0.45$.  The function  $b_1$ takes on its largest value of  about 10$^{-2}$ at the lowest measured value of  $x$ (0.012),   decreases  with increasing $x$  through zero and takes on a minimum value of about $-4\times 10^{-3}$ (with large error bars).

The function $b_1$ vanishes if the spin-one target is made of constituents at rest or in a relative $s$-state, but is very small for a target of spin 1/2 particles moving non-relativistically in higher angular momentum states~\cite{jaffe,miller,Khan:1991qk,Umnikov:1996qv}.  Thus one expects~\cite{jaffe}  that a nuclear $b_1$  may be dominated by non-nucleonic components of  the target nuclear wave function. Consquently, a Jefferson Laboratory experiment~\cite{Jlab} is planned to measure $b_1$ for values of $x$ in the range 
$0.16<x<0.49$  and $1<Q^2<5 $ GeV$^2$ with the aim of reducing the error bars. 

\begin{figure}
\includegraphics[width=4.191cm,height=7.5cm]{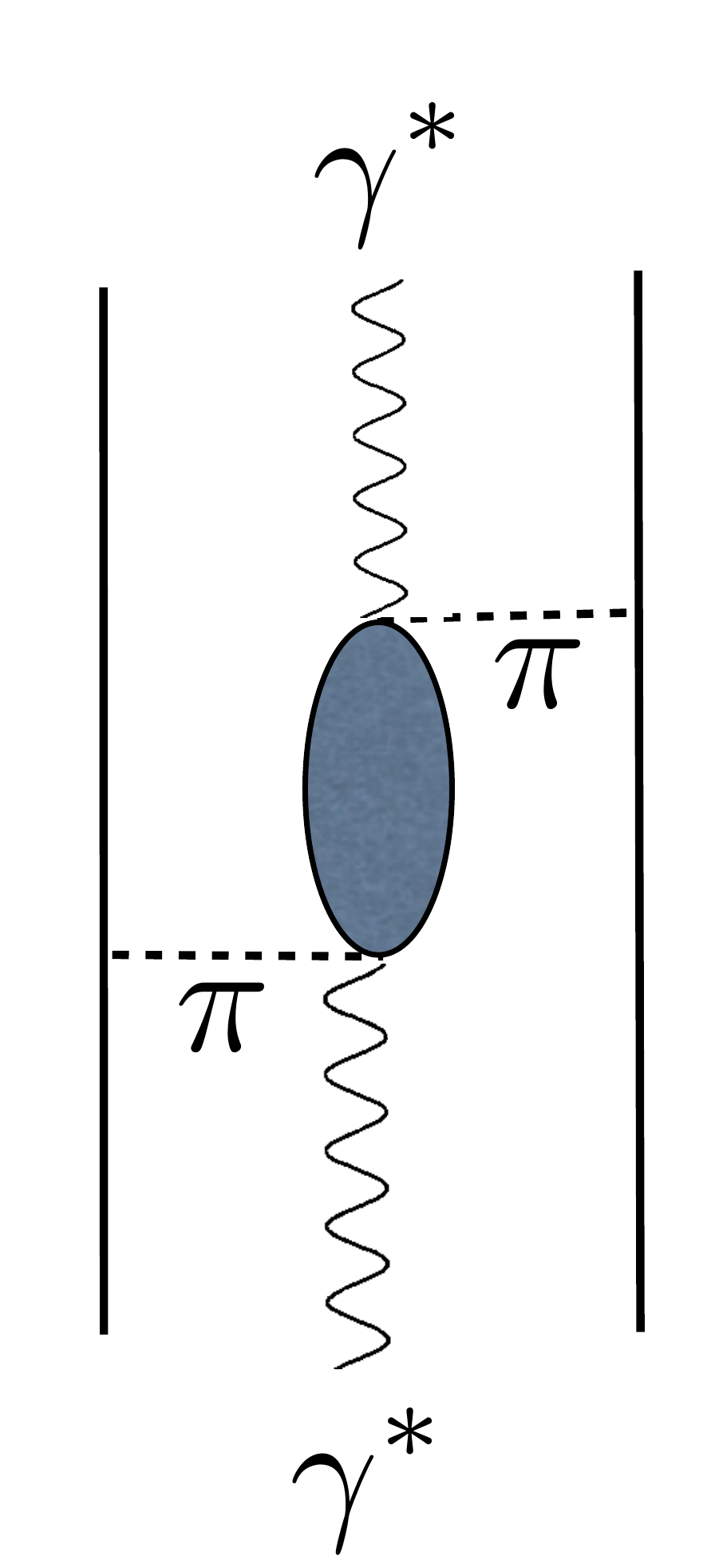}
\caption{Forward Compton scattering diagram for the Sullivan process. The virtual photon $\gamma^*$ encounters an exchanged pion (dashed line), breaking it up
forming a  complicated state (blob) which then emits the pion which is absorbed by another nucleon. The imaginary part of this graph is related to the
deep inelastic structure functions of the deuteron.  }\label{fig:diag}\end{figure}

At very small values of $x$  effects of shadowing (double scattering) are expected to be important~\cite{Nikolaev:1996jy,Edelmann:1997qe,Bora:1997pi}. Our focus here is on the kinematic  region of higher values of $x$ that are available to the JLab experiment.
It is therefore natural to think of the nuclear Sullivan mechanism~\cite{Sullivan:1971kd}, Fig.~\ref{fig:diag}, in which an exchanged pion is struck by  a virtual photon produced by an incoming lepton. That the one-pion exchange potential OPEP gives a tensor force of paramount importance in deuteron physics is a nuclear physics textbook item ~\cite{EW}. Indeed, realistic  deuteron wave functions can be constructed using only the OPEP along with a suitable cutoff at short distances~\cite{Ericson:1981tn, Ericson:1982ei,Friar:1984wi,Cooke:2001rq}.  Therefore it is reasonable to 
estimate the size of such pionic effects. The present author did this in 1989 conference proceeding~\cite{miller}, finding that the effects are small. 
See also~\cite{Nikolaev:1996jy}.
However, as experimental techniques have improved dramatically, the meaning of small has changed. Therefore, considering the planned JLab experiment, it is worthwhile to   re-assess the size and uncertainties of the pionic effects. 

However, the Hermes experiment~\cite{Airapetian:2005cb} presents an interesting puzzle because it observed a significant negative value of $b_1$ for $x=0.45$. At such   a value of $x$, any sea quark effect such as arising from double-scattering or virtual pions is completely negligible. Furthermore, the nucleonic contributions are computed to be very small~\cite{miller,Khan:1991qk,Umnikov:1996qv}, so one must consider other possibilities. We therefore take up the possibility that the deuteron has a six-quark component that is orthogonal to two nucleons.
Such configurations are known to be dominated by the effects of so-called hidden-color states in which two color-octet baryons combine to form a color singlet~\cite{Harvey:1988nk}. In particular,   a component of the deuteron in which all 6 quarks are in the same spatial wave function ($|6q\rangle$) can be expressed in terms on nucleon-nucleon $NN$, Delta-Delta $\Delta\Delta$ and hidden color components $CC$ as~\cite{Harvey:1988nk}:
\bea |6q\rangle=\sqrt{1/9}|N^2\rangle+\sqrt{4/45}|\Delta^2\rangle+\sqrt{4/5}|CC\rangle.\eea
This state has an 80\% probability of hidden color and only an 11\% probability to be a nucleon-nucleon configuration. In the following, the state $|6q\rangle$ is simply referred to as hidden color state.

The discovery of the EMC effect caused researchers to consider the effects of such six-quark states~\cite{Carlson:1983fs} and in a variety of nuclear phenomena~\cite{Miller:1985un,Koch:1985yi,Miller:1987ku}. Furthermore, the possible discovery of  such a state as a di-baryon resonance has drawn recent  interest~\cite{Bashkanov:2013cla}. Therefore we propose a model of a hiden-color six-quark components of the $s$ and $d$-states of the deuteron.  We also note that including a six-quark hidden color component of the deuteron does not lead to a conflict with the measured asymptotic $d$ to $s$ ratio of the deuteron~\cite{Guichon:1983hi}.

Sect. II presents the formalism for computing pionic contributions to $b_1$. Sec. III presents our simple model for the hidden color $s$ and $d$ states of the deuteron.
Sec.~IV  compares the effects of pions and hidden color  with the existing Hermes data and makes predictions for the upcoming JLab experiment. The sum rule of Close \& Kumano~\cite{close} that
$\int dx\,b_1(x)=0$ is discussed in Sec.~V and summary remarks are presented in Sec.~VI.

    \vskip0.2cm\noindent 
   
    \section{One Pion Exchange Effects}
The pionic contribution to the nuclear quark distribution~\cite{Ericson:1983um} for a spin 1 target of $J_z=m$ is given by ~\cite{Jung:1990pu,miller}
\bea \Delta_\pi q^{(m)}(x)=\int_x^\infty{dy\over y}q^\pi(x/y)f^{(m)}_\pi(y),\eea
where $q^\pi(x)$ is the charged-weighted quark structure function of the pion (assumed to be the same for nuclear pions as for free pions):
\bea q^\pi(x)= {5\over 9} u_v^\pi(x) +{10\over9}\bar{u}^\pi+{2\over9 }s^\pi(x),
\eea 
where $u_v^\pi$ is the valence $u$ quark distribution of the $\pi^+$ and $s^\pi$ is the sea quark distribution of a flavor symmetric pion sea,
and 
the probability to find a pion  residing in a deuteron $D$ of $S_z=m$ is given by
\bea
f^{(m)}_\pi(y_A)=\int {d\xi^-\over 2\pi}e^{-i y_A P_D^+\xi^-} \langle D,m|\phi_\pi(\xi^-)\phi_\pi(0)|D,m\rangle_c,\label{feq}
\eea
where the subscript $c$ stands for connected terms.   The matrix element in \eq{feq} is a light-cone correlation function evaluated in the laboratory frame, so that $P_D^+=M_D.$
We suppress the notation for the $Q^2$ dependence of the pion structure function, but include its effects in calculations discussed below.

The resulting  contribution to $b_1$ is given by
\bea b_1^\pi(x)={1\over2}\left(\Delta_\pi q^{(0)}(x)-\Delta_\pi q^{(1)}(x)\right).\eea
The expression,  \eq{feq},  for  $f^{(m)}_\pi$ is evaluated by saturating the intermediate states with 2 nucleon,~1~pion states.  We use the nucleon variable $y$ with
$y M= y_A M_D$ ($M$ is the nucleon mass).  Evaluation using non-relativistic dynamics and neglecting retardation effects in the pion propagator leads to
\bea
f^{(m)}_\pi(y)={-3y g^2\over (2\pi)^3}\int {d^3q\over \left(\bfq^2+m_\pi^2\right)^2}{G_A^2(\bfq^2)\over G_A^2(0)} \delta(M y-q_z)F_m(\bfq),\label{fm}\eea
with 
\bea
F_m(\bfq)\equiv\int d^3r\langle D,m| e^{-i\bfq\cdot\bfr}\boldsigma_1\cdot\bfq\,\boldsigma_2\cdot\bfq|D,m\rangle, \label{Fm}\eea
where $\bfr$ is the displacement between the neutron and proton, $g$ is the pion-nucleon coupling constant (we use 13.5) and $M$ is the nucleon mass. The nucleons involved in   non-relativistic nuclear wave functions are on their mass-shell. This means that one may use 
   the 
generalized Goldberger-Treiman relation~\cite{Thomas:2001kw}) to relate the pion-nucleon form factor $ G_{\pi N}(t)$  to the axial  form factor:
 \bea  G_{\pi N}(t)={M\over f_\pi}G_A(t),\label{related}\eea where $t$ is the square of the four-momentum transferred to the 
 nucleons, $G_A(t)$ is the axial vector form factor and  $f_\pi$ is the pion decay constant and $G_{\pi N}(0)\approx g$.  Using  \eq{related} has obvious practical value because it relates an essentially unmeasurable quantity $G_{\pi N}$ with one $G_A$ that is constrained by experiments, and has been used in obtaining \eq{fm}.
We use the dipole form:
  $ G_A(Q^2)/G_A(0)=1/(1+(Q^2/M_A^2))^2,$
  with $M_A$ as the so-called axial mass. 
  The values of $M_A$ are given by $M_A=1.03\pm 0.04$ GeV as reviewed in \cite{Thomas:2001kw}. This range is consistent with the one reported in a later review~\cite{Bernard:2001rs}.  
 
   \begin{figure}[t]
\includegraphics[width=7.091cm,height=6.5cm]{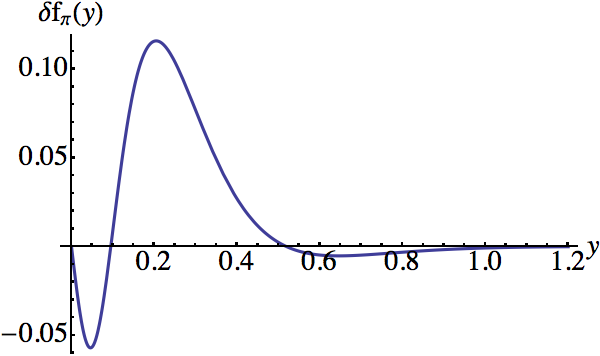}
\caption{$\delta f_\pi(y)$ of \eq{dfyd}. The results obtained with the Argonne V18 deuteron wave function~\cite{Wiringa:1994wb} overlap with  those of the Reid '93 potential~\cite{Stoks:1994wp}.}\label{dry}\end{figure}

   To proceed it is convenient to use the following representation~\cite{EW}  of the deuteron wave function:
   \bea\langle\bfr|D,m\rangle={1\over\sqrt{4\pi}}\left[{u(r)\over r}|1,m\rangle+{w(r)\over r}{(3\boldsigma_1\cdot \hat{\bfr}\boldsigma_2\cdot\hat{\bfr}-1)\over\sqrt{8}}|1,m\rangle\right],\eea
   where $|1,m\rangle$ represents the triplet spin wave function. Evaluation of \eq{fm} yields the results
   \bea F_m(\bfq)=F_m^{uu}(\bfq)+F_m^{uw}(\bfq)+F_m^{ww}(\bfq),\eea
   where 
   \bea& &F^{uu}_{\pm1}=q_z^2I_{uu0}(q),\,F^{uu}_{0}=(q_\perp^2-q_z^2)I_{uu0}(q),\\
&& F^{uw}_{\pm1}=-{1\over\sqrt{2}}(3\bfq^2-q_z^2)I_{uw2}(q),\,F^{uw}_{0}=-{1\over\sqrt{2}}(3\bfq^2-\left(q_\perp^2-q_z^2)\right)I_{uw2}(q),\\  
&& F^{ww}_{\pm1}=q_z^2I_{ww0}(q)+{1\over4}(3\bfq^2-q_z^2)I_{ww2}(q),\,\\&&
 F^{ww}_{0}=(q_\perp^2-q_z^2)I_{ww2}(q)+{1\over4}\left(3\bfq^2-(q_\perp^2-q_z^2)\right)I_{ww2}(q),\\
 &&I_{abL}(q)\equiv \int_0^\infty dr\, a(r)b(r)j_L(qr).
  \eea
  where $a,b=u,w$, $L=0,2$ and $\bfq^2=q^2=q_z^2+q_\perp^2$.
We need the combinations $F_0^{ab}(\bfq)-  F_1^{ab}(\bfq)\propto(\bfq^2-3q_z^2)$  to compute $b_1$. Therefore it is useful to define the integral which gives the individual terms of $f_\pi^{(0)}(y)-f_\pi^{(1)}(y):$
\bea&& f_{abL}(y)\equiv-{3yg^2\over(2\pi)^2}\int {d^3q\over (\bfq^2+m_\pi^2)^2}{\delta(My-q_z)\over(1+{\bfq^2\over M_A^2})^4}\left(\bfq^2-3q_z^2\right)I_{abL}(q),\label{cool}\\
&&=-{3yg^2\over8\pi^2}\int_0^\infty {dq_\perp^2\over(q_\perp^2+M^2y^2+m_\pi^2)^2}{1\over(1+{q_\perp^2+M^2y^2\over M_A^2})^4}\left(q_\perp^2-2M^2y^2\right)I_{abL}(\sqrt{q_\perp^2+M^2y^2}).
\eea
 Then
 \bea \delta f_\pi(y)\equiv f_\pi^{(0)}(y)-f_\pi^{(1)}(y)=f_{uu0}(y)+{\sqrt{2}\over2}f_{uw2}(y)+f_{ww0}(y)-{1\over4}f_{ww2}(y),\label{dfyd}\eea
 and
 \bea  b_1^\pi(x)={1\over2}\int_x^\infty{dy\over y}q^\pi(x/y)  \delta f_\pi(y).\label{bpif}\eea
  
 The key output of the present section is  the function $\delta f_\pi(y)$, which is displayed in  Fig.\ref{dry}. 
The Argonne V18 deuteron wave function ~\cite{Wiringa:1994wb} is used here, but virtually  identical results are obtained with the Reid '93 potential~\cite{Stoks:1994wp}. Note the double node structure, a consequence of the tensor nature of the operator, that can be understood by examining \eq{cool} and the functions $I_{abL}(q).$
For small values of $y$, $f_{abL}(y)\propto (-y)$, but for larger values of $y=q_z$, the integrand changes sign.  A node in the functions $I_{abL}(q)$ causes another sign change  at still larger values of $y$. Indeed, we may use \eq{cool} to obtain a sum rule:
\bea \int_{-\infty}^\infty dy{f_{abL}(y)\over y}=2 \int_{0}^\infty dy{f_{abL}(y)\over y}=0,\label{msr}\eea with the 0 resulting from the feature $\int d^3q f(\bfq^2)(\bfq^2-3q_z^2)=0$.
This sum rule has been used as a numerical check on the integrals.  No general result for $\int dy\,f(y)$ can be obtained because of the factor $y$ appearing in front of the integral in \eq{cool}.

\section{Hidden-color six-quark states} 
  
 We  investigate the  possible relevance of hidden-color six-quark states. For this purpose, it is sufficient to use the simplest of many possible models. Thus we assume a deuteron component  consisting of
  six non-relativistic quarks in an $S$-state. As stated above, such a state has only a probability of 1/9 to be a nucleon-nucleon component, and is to a reasonable approximation a hidden color state, so we use the terminology six-quark, hidden color state.  Then we obtain  the corresponding  $d$ state by  promoting any one of the quarks to a $d_{3/2}$-state. We define these states by combining
  5 $s$-state quarks into a spin 1/2 component, which couples with the either the $s_{1/2}$ of $d_{3/2}$ single-quark state to make a total angular momentum of 1.
 We therefore write the  wave functions of these states for a deuteron  of $J_z=H$  as
  
 \bea
  \psi_{j,l,H}(\bfp)=\sqrt{N_l}f_l(p)\sum_{m_s,m_j}{\cal Y}_{jlm_j}\langle jm_j,{1\over2}m_s\vert1H\rangle,\label{wf}
 \eea 
 where $l,j=s_{1/2}$ or $d_{3/2}$,  
  $N_l$ is a normalization constant chosen so that $\int\,d^3p\bar{\psi}_{j,l,H}(\bfp)\gamma^+\psi_{j,l,H}(\bfp)=1$ and ${\cal Y}_{jlm_j}$ is a spinor spherical harmonic. The  matrix element  for transition between the $l=0$ and $l=2$ states is given by the light-cone distribution:
\bea &F_H(x_{6q})={1\over2}\int d^3p\bar{\psi}_{1/2,0,H}(\bfp)\gamma^+\psi_{3/2,2,H}(\bfp)\delta\left({p\cos\theta+E(p)\over M_{6q}}-x_{6q}\right),
\eea
where $E(p)=\sqrt{p^2+m^2}$ with $m$ as the quark mass, and $M_{6q}$ is the mass of the six-quark bag, $x_{6q}$ is the momentum fraction of the six-quark bag carried by a single quark and $x_{6q}M_{6q}=x M$~\cite{Carlson:1983fs}. Note that
  $p\cos\theta$ is the third ($z$)  component of the momentum, so that the plus component of the quark momentum is $E(p)+p\cos{\theta}$. 
   We take $M_{6q}=2M$ (its lowest possible value) to make a conservative estimate.
  
 The term of interest $b_1(x)$ is given by
 \bea b_1^{6q}(x)={1\over2}(2)\left(F_0(x)-F_1(x)\right)P_{6q},\eea
 where $P_{6q}$  is the product of the  probability amplitudes for the 6-quark states to exist in the deuteron, and the factor of 2 enters because either state can be in the $d$-wave.
 Evaluation of $F_H$ using \eq{wf} leads to
 the result:
 \bea& b_1^{6q}(x)=-\sqrt{N_0N_2\over 2}{3\over4\pi}\int d^3p f_0f_2 (3\cos^2\theta-1)\delta\left({p\cos\theta+E(p)\over M}-x\right)P_{6q}.\nonumber\\&\label{b16}\eea
To proceed further we   specify the wave functions to be  harmonic oscillator wave functions. We  take $f_2(p)=-p^2R^2e^{-p^2R^2/2},\, f_0(p)=e^{-p^2R^2/2},$ where $R$ is the radius parameter.   $R$ is chosen as 1.2  fm, which corresponds to the measured radius of the nucleon and the notion that the bag model stability condition gives the radius of the 6-quark bag to be about 1.4 times the nucleon radius.  We use a quark mass of 338 MeV~\cite{Close:1979bt}. A spread of values of the model parameters $R,m$  around the central values of $1.2\,{\rm fm},\,338 $ MeV
will be examined below.

The evaluation of $b_1^{6q}(x)$ proceeds by using $d^3p=2\pi p^2dpd\cos\theta $, integrating over $\cos\theta$,  and changing variables to $u\equiv\,p^2R^2$.
The result is
\bea b_1^{6q}(x)={6MR \over\sqrt{30\pi}}\int_{u_{\rm min}(x)}^\infty\,du\,e^{-u}\left[3((x^2M^2+m^2)R^2+u-2xM R\sqrt{u+m^2R^2})-u\right]P_{6q},\label{b6q}\eea
where
\bea u_{{\rm min}(x)}\equiv {(x^2M^2-m^2)^2R^2\over 4x^2M^2}.\eea

        \begin{figure}
\includegraphics[width=10.991cm,height=8cm]{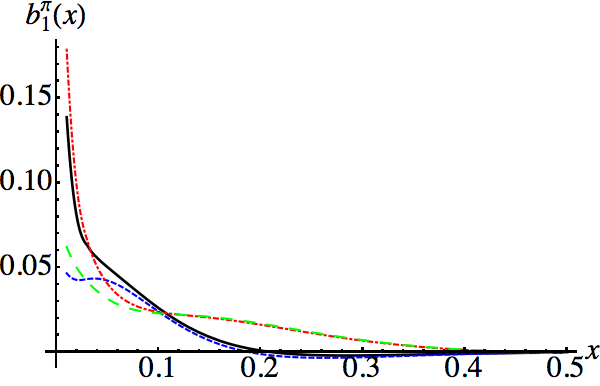}
\caption{Color online. Computed values of $b_1^\pi$, for different pion structure  function at $Q^2=1.17$ GeV$^2$. Solid- full structure function~\cite{Aicher:2010cb}\,  short-dashed (blue)  valence~\cite{Aicher:2010cb},  Dot Dashed (Red) full structure function (mode 3)~\cite{Sutton:1991ay},Long dashed (green)   (mode 3)~\cite{Sutton:1991ay} } \label{fig:SFPlot}\end{figure}

 \section{Results}

  \begin{figure}
\includegraphics[width=8.991cm,height=10cm]{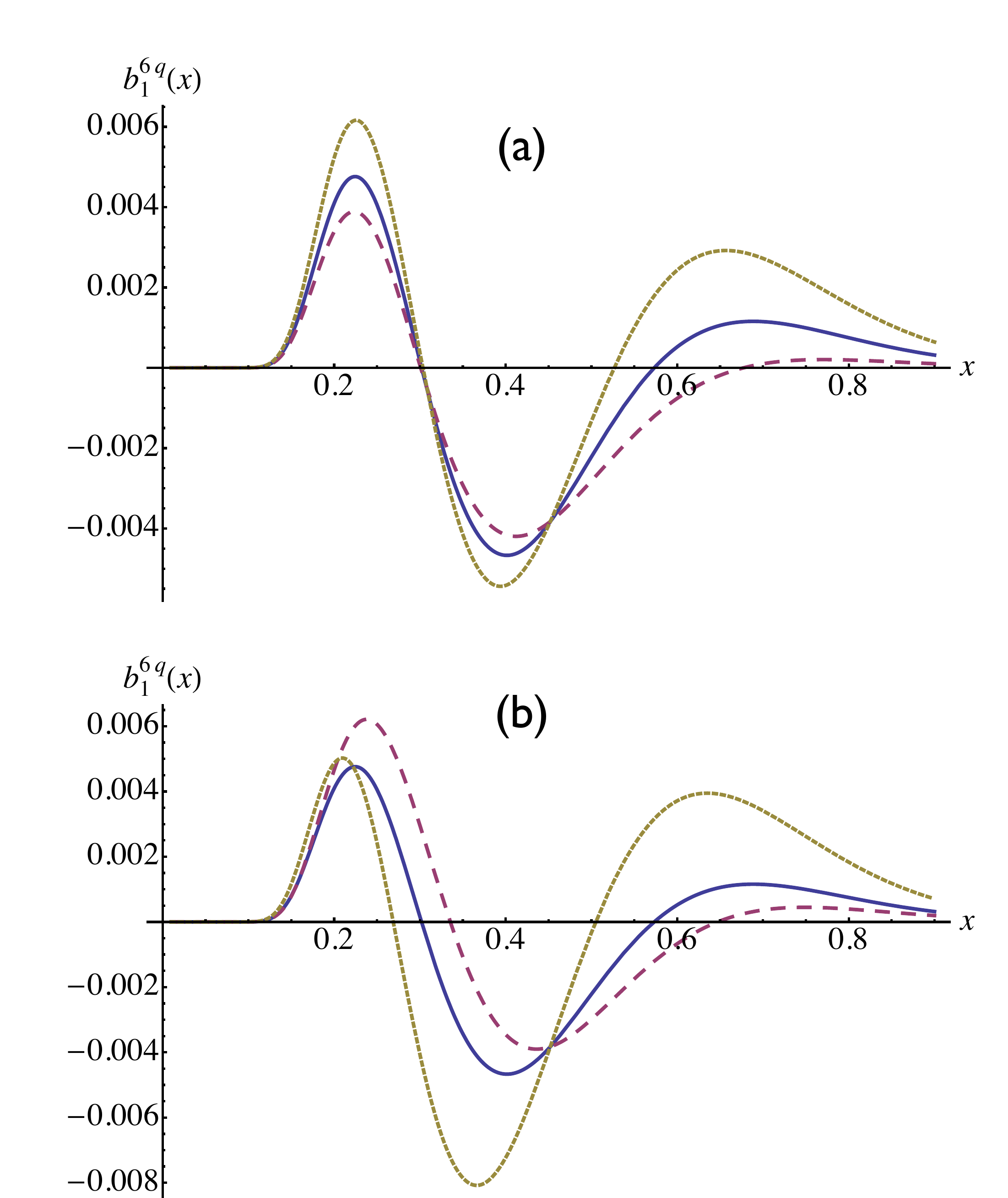}
\caption{(Color online) Computed values of $b_1^{6q}$ from \eq{b6q}. Sensitivity to parameters is displayed.  (a) Solid (blue) uses $R=1.2$ fm, m=338 MeV, long dashed (Red) $R$ is decreased by 10\%, dotted(green) $R$ is increased  by 10\%. (b) Solid (blue) uses $R=1.2$ fm, m=338 MeV, long dashed (Red) $m$ isincreased by 10\%, dotted(green), $m$ is decreased  by 
10\%.} \label{fig:6q}\end{figure}

We may now start examining the resulting phenomenology, considering first the pionic contributions.
The   quark distribution function of the pion, $q^\pi$  is  needed to evaluate $b_1^\pi$ as shown in \eq{bpif}. Evaluation  requires knowledge of this function over a wide range of its argument and $x/y$ can be very small.
However, knowledge of  $q^\pi$ comes from fixed-target Drell-Yan data at values of $x\ge 0.3$~\cite{Conway:1989fs,Bordalo:1987cr}.
We display the sensitivity to different versions of $q^\pi$ in Fig.~\ref{fig:SFPlot}.  We display  results for the  full and valence distributions of~\cite{Aicher:2010cb}, and
for the full and valence distributions of ~\cite{Sutton:1991ay} at $Q^2=1.17$ GeV$^2$.
The sea is important for values of $x$ less than about 0.1.  This is unfortunate because the Drell-Yan data at large $x$ embody  little sensitivity to the sea. However, the computed values of $b_1^\pi$ are not very different for the two parameterizations, except for very small values of $x$. 
 The solid and dashed curves show the result of using
two different valence quark distributions of~\cite{Aicher:2010cb}  at $Q^2=0.4 $ GeV$^2$. For Fit 3 (solid) $q^\pi(x)\sim x^{-0.3} $ while for Fit 4 (dashed)  $q^\pi(x)\sim x^{0.06} $ .

We now turn to the determine the contributions due to hidden color, $b_1^{6q}$ provided  by \eq{b6q}.
The value of $b_1^{6q}$  $x=0.452$  is relevant because the pionic contribution is negligible, and  the measured value, $b_1=-3.8 \pm 0.16\times 10^{-3}$, differs from zero.
We choose  $P_{6q}=0.0015$ to reproduce the central value using $R=1.2$ fm and $m=338 $ MeV.  Such a very, very small value can not be ruled out by any observations.

The results for $b_1^{6q}$
 are shown in Fig.~\ref{fig:6q}.  Results using the model  parameters $R=1.2$ fm, $m=338 $ MeV are shown as the solid curves in  Figs.~\ref{fig:6q}a and b.
 The exponential appearing in \eq{b6q} renders the contribution very small for small values of $x$. This is seen in the figure. However, there is a large negative contribution at values of $x\approx0.4$, as well as a double-node structure. The latter arises from the factor $3\cos^2\theta-1$ appearing in the integrand of \eq{b16}. The contributions of the hidden-color configurations are generally much smaller than those of exchanged pions except for values of $x$ larger than about 0.35. We also predict that, for even larger values of $x$,  $b_1$ changes sign and may have  another maximum. This mechanism allows contributions at large values of $x$. A quark in a hidden color, six quark configuration can have up to two units of $x$.
 The parameter dependence of the model is also explored. Fig.~\ref{fig:6q}a shows the dependence on the value of $R$ and 
Fig.~\ref{fig:6q}a shows the dependence on the value of $m$.  For each of the curves $P_{6q}$ is chosen so that the value at $x=0.452$ is the same. Shifting the value of $R$ while keeping $b_1^{6q}(0.452)$ fixed  requires less than 4\% changes in the value of $P_{6q}$. Increasing the value of the quark mass produces larger effects. Keeping $b_1^{6q}(0.452)$ fixed   requires that the value of
 $P_{6q}$ needs to be decreased by 20\% if the value of the quark  mass is increased by 10\%, and the value of  $P_{6q}$ needs to be increased by about a factor of 1.8 if the value of the quark  mass is decreased by 10\%. In the remainder of this paper, we use the central values $R=1.2$ fm, $m=338 $ MeV.

\begin{figure}
  \includegraphics[width=10.991cm,height=8cm]{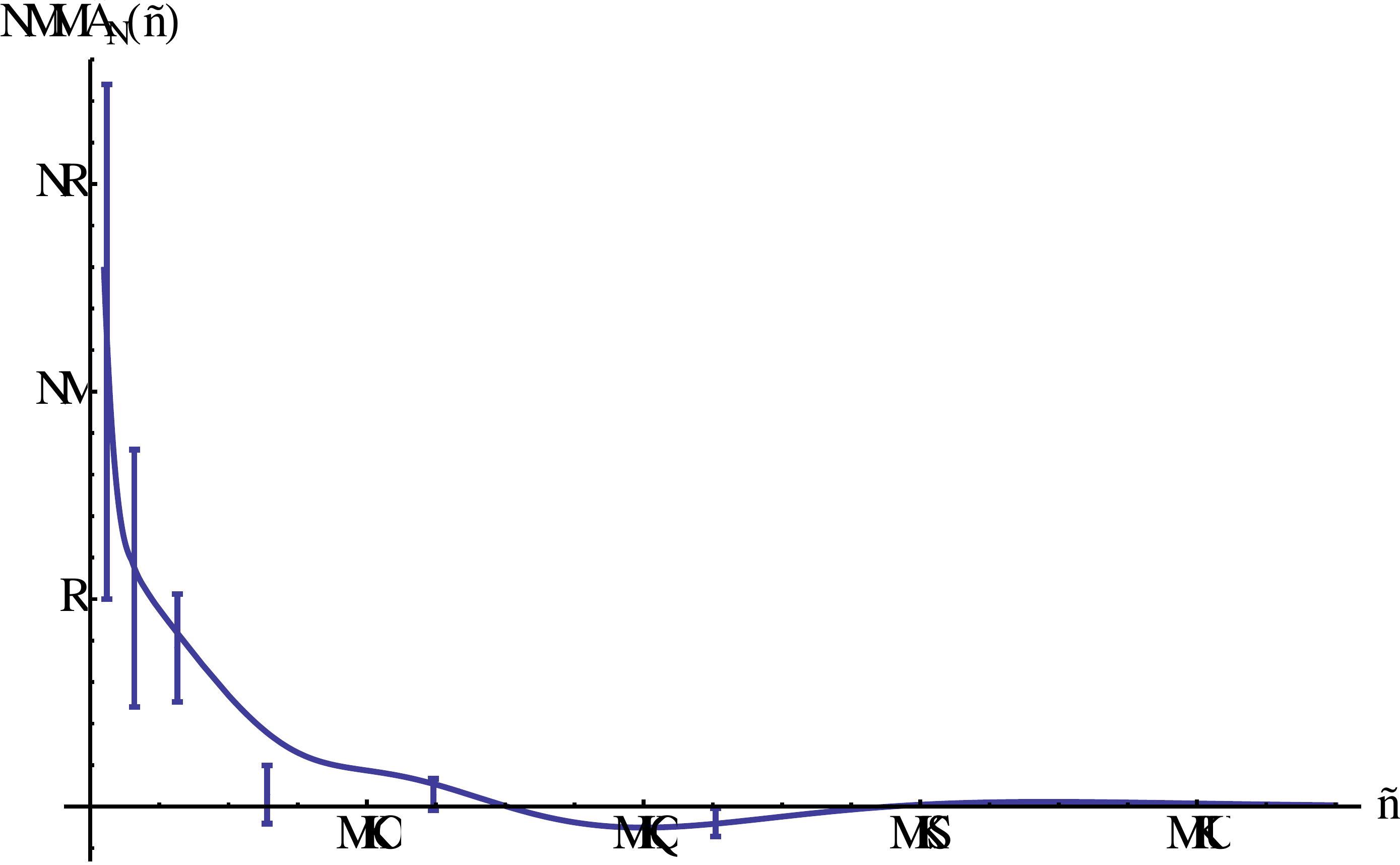}
\caption{ Computed values of $b_1=b_1^\pi+b_1^{6q}$ from \eq{bpif} and \eq{b6q}. The pion structure function is that of ~\cite{Aicher:2010cb},  model 1}\label{HD}\end{figure}

 At this stage we can assess the size of our  computed $b_1^\pi$ and $b_1^{6q}$ versus the only existing data~\cite{Airapetian:2005cb}. These data is given in Table~\ref{b1result} along with our computed values of $b_1^\pi$ using the pion structure functions of Re.~\cite{Aicher:2010cb} and the three modes of~\cite{Sutton:1991ay}. These modes differ in the fraction of momentum carried by the sea: 10\%,15\% and 20\% for modes 1,and 3 respectively. The differences obtained by using different structure functions are generally not larger than the experimental error bars.
For values of  $x$ less than about  $0.2$, there is qualitative agreement between the measurements and the calculations of $b_1^\pi$ (which are much larger than those of $b_1^{6q}$), given  the stated  experimental uncertainties  and  the unquantifiable uncertainty caused by lack of knowledge of the sea. However, the large-magnitude negative central value measured at $x=0.452$ is two standard deviations away from the value provided by $b_1^{\pi}$ but in accord with the value provided by $b_1^{6q}$. 
Thus our result is that one can reproduce the Hermes measurements by using pion exchange contributions at low values of $x$ and hidden-color configurations at  larger values of $x$. This is also shown in Fig.~\ref{HD}, where very good agreement between data and our model can be observed.
 The contributions  of double scattering~\cite{Bora:1997pi} are far smaller than the measurements for the values of $x$  displayed in the table and in Fig.~\ref{HD}, and are therefore neglected here.

\begin{figure}
\includegraphics[width=10.991cm,height=8cm]{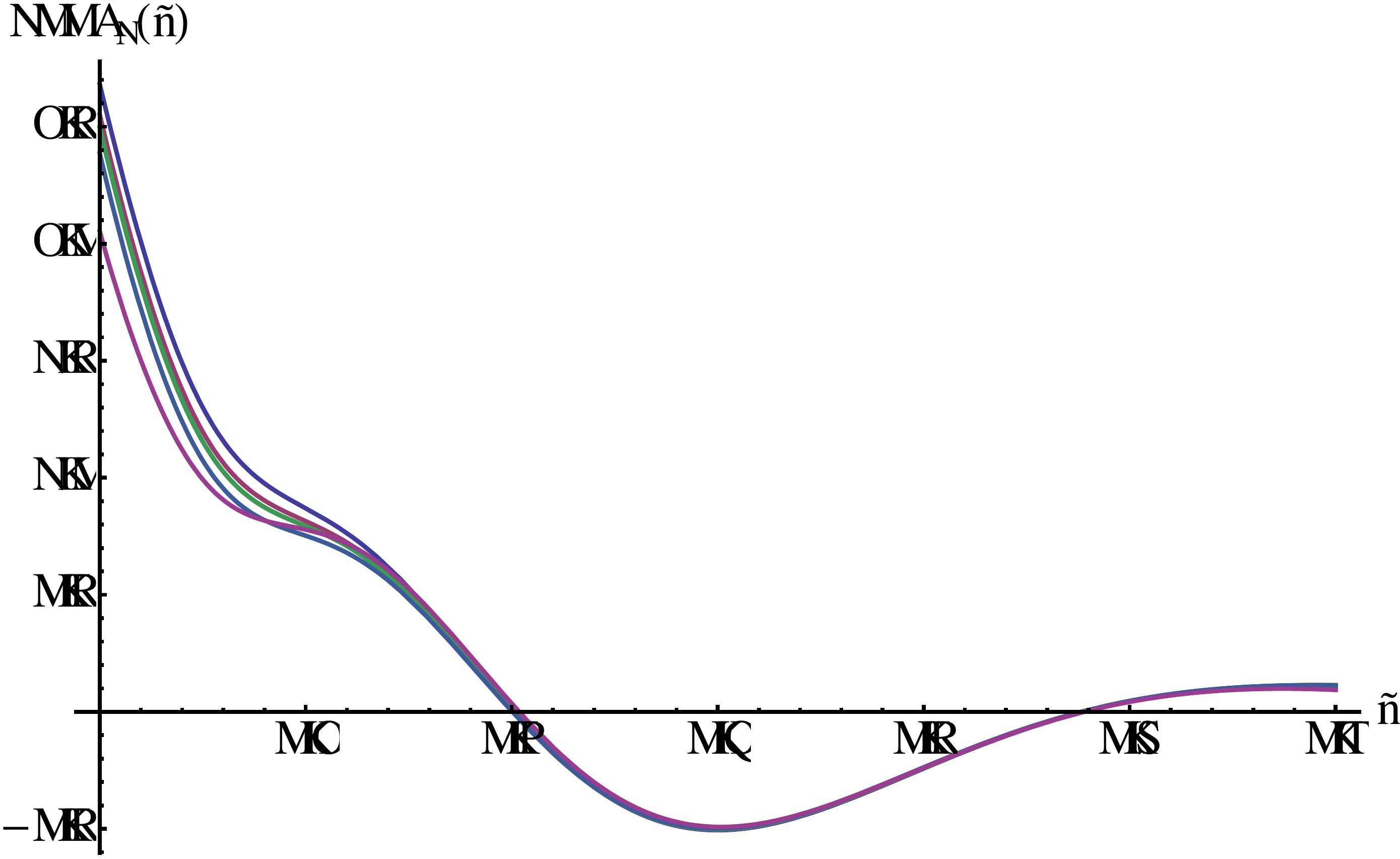}
\caption{(Color online) Computed values of $100\,(b_1^\pi+b_1^{6q})$, for  values of $Q^2\,=\,$ 1.17, 1.76, 2.12 and 3.25 GeV$^2$~\cite{Aicher:2010cb} distributions and for ~\cite{Sutton:1991ay} (lowest curve at $x=0.15$). For the other curves,   $b^\pi_1$ increases as $Q^2$ increases for small values of $x$.   } \label{JLab}\end{figure}

 \def\vum#1{\vspace*{#1cm}}
 \def\jum#1{\hspace*{#1cm}}
\begin{table}[b]  
\caption{ \label{b1result}
Measured values (in $10^{-2}$ units) of the  tensor structure function $b_1$. 
Both the  statistical and systematic uncertainties are listed.  The numbers in parenthesis refer to the structure function modes of Ref.~\cite{Sutton:1991ay} .}
\vum{-0.0}
\begin{tabular}{ccrrrrrrr} \hline\hline
$\langle x \rangle$ & $\langle Q^2\rangle$ &  
 $b_1$  &   $\pm\delta {b_1}^{\!\rm stat}$ &   $\pm\delta {b_1}^{\!\rm sys}$ & $\jum1 b_1^\pi$\cite{Aicher:2010cb}&$b_1^\pi$\cite{Sutton:1991ay} (1)&$b_1^\pi$\cite{Sutton:1991ay} (3) & $\quad b_1^{6q}$ \\
   & $\rm [GeV^2]$ &    $[10^{-2}]$ & $[10^{-2}]$ & $[10^{-2}]$ &\,\quad\hspace{.5cm}  $[10^{-2}]$ & $[10^{-2}]$  & $[10^{-2}]$&$\quad[10^{-2}]$\\
\hline
  0.012 &  0.51 &   11.20 &   5.51 &   2.77 &10.5 &15.5&24.1 &\quad0.00 \\
  0.032 &  1.06 &     5.50 &   2.53 &   1.84 &5.6 &6.8&8.9&0.00 \\
  0.063 &  1.65 &    3.82 &   1.11 &   0.60  &  4.2&3.7&4.1&0.00\\
  0.128 &  2.33 &      0.29 &   0.53 &   0.44 & 1.6& 1.3&1.3&0.01\\
  0.248 &  3.11 &    0.29 &   0.28 &   0.24 &-0.55& .13&0.12&0.41\\
  0.452 &  4.69 &    -0.38 &   0.16 &   0.03 &-0.02&-0.02&-0.022&-0.38\\ \hline\hline
\end{tabular}
\vum{-0.0}
\end{table}
The next step is to make predictions for the JLab experiment.
Our results for $b_1= b_1^\pi+ b_1^{6q}$ are shown in Fig.~\ref{JLab}. For values of $x$ less than about 0.2 the computed values of $b_1$ are dominated by those of  $b_1^\pi$.   For larger values of $x$ the computed results are not significantly different from 0. This result combined with the very small large $x$ results for  nucleonic~\cite{jaffe,Khan:1991qk,Umnikov:1996qv} and double-scattering  contributions~\cite{Nikolaev:1996jy,Edelmann:1997qe,Bora:1997pi}, makes the case that an observation of a value of $b_1$  significantly different  than  zero for values of  $x$ greater than about 0.3 would represent a discovery of some sort of exotic nuclear physics. Our model \eq{b6q} leads to an effect that does contribute at larger values of $x$. One may roughly think of the prediction for the JLab experiment as arising from $b_1^\pi$ for $x<0.2$ and from $b_1^{6q}$ for $x>0.2$.

\section{Sum Rule of Close \& Kumano~\cite{close}}
Close \& Kumano found a sum rule that the integral of $b_1(x)$ vanishes:
\bea \int dx b_1(x)=0, \label{cksr}\eea
provided that the sea is unpolarized, as is the case for the pion contribution discussed here.
This sum rule is interesting because it shows that if $b_1(x)$ is significantly different from 0 at one value of $x$, it must take on significant values of the opposing sign for other values of $x$.

A visual inspection of the Fig.~\ref{fig:SFPlot} shows immediately  that the pionic contribution does not obey this sum rule. This result can be seen analytically by integrating \eq{bpif} over $x$:
\bea&& \int_0^1 dx b_1^\pi(x)={1\over2}\int_0^1\,dx\int_x^\infty{dy\over y}q^\pi(x/y)  \delta f_\pi(y)\\
&&={1\over2}\int_0^2dy  \delta f_\pi(y)\int_0^1 du q^\pi(u).\eea
The above result is obtained by interchanging the  order of the integration over $x$ and $y$ and changing variables 
from $x$ to $u=x/y$.
There are two reasons why the product of integrals on the right does not vanish. The first is displayed above in \eq{msr};
the integral of  $\delta f_\pi(y)/y$ vanishes, so the integral of $\delta f_\pi(y)$ can not vanish. The second is that the integral of the quark distribution function of the pion is infinite for any of the published structure functions. Thus the value of the sum rule is infinity.

Given this violation of the sum rule of Close \& Kumano, it is interesting to see if the extant calculations of other mechanisms are consistent with the sum rule. Consider first the nucleonic 
contribution~\cite{jaffe,Khan:1991qk,Umnikov:1996qv}. In particular we examine  Eq.~(15,16) of \cite{Khan:1991qk}:
\bea &&b_1^N(x)=\int_x^2{dy\over y}\Delta b(y)F_{1N}(x/y)\\ 
&&\Delta b(y)=\int d^3p F_d(p)(3 \cos^2\theta-1)(1+{p\cos\theta\over M}+{p^2\over 4M^2})\delta\left(y-{p\cos\theta+E(p)\over M}\right)\\
&&F(p)\equiv -{3\over 4\pi\sqrt{2} }\sin\alpha\cos\alpha u_s(p)u_d(p)+{3\over 16\pi}\sin^2\alpha u_d^2(p),
\eea
where $M$ is the nucleon mass, $E(p)=\sqrt{p^2+M^2}$, and $u_s,u_d$ are the $s$ and $d$ state components of the deuteron wave function.  The angle $\theta$ is the polar angle for the vector ${\bfp}$.  Integration over $x$  and repeating the above manipulations leads to the result:
\bea \int_0^1 dx b_1^N(x)=\int_0^2dy\Delta b(y)\int_0^1 du F_{1N}(u).\eea
There is  again a  product of integrals, with the one of $F_{1N}$ being infinite for any reasonable structure function.
Close \& Kumano state that the integral over $y$ vanishes. If that were correct the value of the sum rule would be zero times infinity, or indefinite. However, while the integral over $y$ is indeed very small, we can show that  it is positive definite. 
First note that
\bea \int_0^2\delta(y-{(E-p\cos\theta)\over M})=\theta(2M- E+p\cos\theta).\label{thf}\eea
If the right-hand-side were unity, the integral of $\Delta b(y)$ would indeed vanish. Using \eq{thf} we find 
\bea&&\int_0^2dy\,\Delta  b(y)=2\pi \int_{-1}^1dz (3z^2-1)\{\theta(z)\int_0^{M\gamma_+(z) }p^2dp F(p)(1+{p^2\over4M}+{p\over M}z)+\nonumber\\&&+\theta(-z)\int_0^{M\gamma_-(z) }p^2dp F(p)(1+{p^2\over4M}+{p\over M}z)\}
\\&&\gamma_\pm(z)\equiv \sqrt{{\sqrt{3} \over1-z^2}+{4 z^2\over(1-z^2)^2}}\mp z\eea
Algebraiic manipulation leads to
\bea
&&\int_0^2dy\,\Delta  b(y)=4\pi \int_{0}^1dz (3z^2-1)\int_0^{M\gamma_+(z) }p^2dp F(p)(1+{p^2\over4M})
\\&&=4\pi \int_{0}^1dz (z-z^3) {(M\gamma_+(z) })^2 F(M\gamma_+(z) )(1+{\gamma_+(z) ^2}),\eea
with the last step obtained by integration by parts.
 The function  $F(p)$ is determined from deuteron wave functions and $F(p)>0$ for values of $p$ less than about 2.5 fm$^{-1}$. For larger momenta $F(p)$ is very small so that
 the integrand is positive  for the important values. This means that the  sum rule takes on the value  infinity.

  \newcommand{\xq}{(x,Q^2)}
Another known mechanism is double scattering, $b_2^{(2)}\xq\ $. Here we use the result of \cite{Bora:1997pi}, obtained after integrating over $k_\perp$ in their Eq.~(20):
 \begin{eqnarray}
b_2^{(2)}\xq\ & = & \frac{-3}{(\pi)^4}\frac{Q^2}{16\sqrt{2}\alpha}\,{\rm Im}\, i
\int d^2  b\,
\int dz \, u_0(r)u_2(r)
\frac{2z^2-b^2}{(z^2+b^2)^2}\times\nonumber\\
 & \times &{\pi\over a}e^{-{b^2\over 4a}}
\sum_V  e^{iz/\lambda_V} 
\frac{M_V^4}{(M_V^2+Q^2)^2} \left.\frac{d\sigma}{dt}\right|_{\gamma N
\rightarrow VN,t=0}, 
\label{looks1}
\end{eqnarray}
where
 \begin{equation}
\lambda_V = \frac{2\nu}{M_V^2+Q^2} = \frac{Q^2}{Mx(M_V^2+Q^2)}\equiv {\Lambda_V\over x}.
\label{coher}
\end{equation}
The $x$ dependence enters through the dependence of $\lambda_V$ on $x$.

To test the sum rule we integrate over $x$.
 \begin{eqnarray}
\int_0^1\,dx b_2^{(2)}\xq\ & = & \frac{-3}{(\pi)^4}\frac{Q^2}{16\sqrt{2}\alpha}\,
\int d^2  b\,
\int dz \, u_0(r)u_2(r)
\frac{2z^2-b^2}{(z^2+b^2)^2}\times\nonumber\\
 & \times &{\pi\over a}e^{-{b^2\over 4a}}
\sum_V {\Lambda_V \over z} 
\sin {z\over \Lambda_V}
\frac{M_V^4}{(M_V^2+Q^2)^2} \left.\frac{d\sigma}{dt}\right|_{\gamma N
\rightarrow VN,t=0}.
\label{looks2}
\end{eqnarray}
The three-dimensional spatial integral involving $(2z^2-b^2)=3z^2-r^2$ would vanish if multiplied by a function  that depended only on $r$. However, the integrand contains  the exponential involving $b^2$ and the $\sin$ involving $z$.
Therefore the integral does not vanish and the  double scattering term violates the sum rule.  The non-vanishing of the sum rule integral for the pionic and double scattering mechanisms   is  in agreement with an earlier finding by~\cite{Nikolaev:1996jy}.

Thus three published mechanisms that  contribute to $b_1$, and all violate the sum rule of Close \& Kumano.
However, one can see from \eq{b16} that  \bea \int dx b_1^{6q}(x)=0.\eea
The integral over all values of $x$ leads to a three-dimensional integral involving $3\cos^2\theta-1$  which must vanish. Moreover, a glance at Fig.~\ref{fig:6q} leads to the expectation that the integral of $b_1^{6q}(x)$ over the values of $x$ displayed vanishes. Indeed, numerical integration leads to a zero within one part in 10$^8$. Furthermore, 
the model of~\cite{jaffe}  involving massless relativistic quarks with $j=3/2$ moving in a central potential also satisfies the sum rule for the same reasons.

Given the different  possible values that the integral of $b_1$ may take,  it seems reasonable to re-examine the derivation and meaning of the sum rule of \eq{cksr}. The key equations  of that paper are their Eq. (15,16):
\bea &&
\Gamma_{H,H}=\langle p,H|J_0(0)|p,H\rangle,\label{relate}\\&&
{1\over2}(\Gamma_{0,0}-\Gamma_{1,1})=\sum_i e_i\int dx \delta q_{i,v}^A(x)\label{parton}\eea 
where
the sum is over quarks of  flavor $i$ and charges $e_i$ of a hadron of $z$-projection $H$. The term $\delta q_i$ is defined in \eq{pm1}, and involves only valence quarks,which have non-zero integrals over all $x$ that are  related to baryon number and charge.

That the sum rule does not hold for mechanisms  in which $b_1$ is generated by non-valence quark contributions, such as the double-scattering and pion exchange mechanisms is consistent with the derivation based on valence quark dominance. The sum rule does not hold for the contributions of  the $d$ state nucleons, but those contributions to $b_1$ are nearly vanishing for non-zero values of $x$. The sum rule does hold for the hidden-color six-quark configurations, in which a valence quark  contributes. Thus the sum rule is a useful guide to the physics relevant for $b_1$. An observation of its failure means that sea effects are important. Measuring significant positive and negative values of $b_1$ at large $x$ could  signify the importance of an exotic valence quark effect.
  
 \section{Summary}
 This paper contains an evaluation of the pion exchange and six-quark, hidden-color contribution to the $b_1$ structure function of the deuteron. The pion-nucleon form factor is constrained phenomenologically to 
 reduce a possible uncertainty. There is some numerical sensitivity to using different pionic structure functions. The pionic mechanism is sizable for small values of $x$, and can reproduce Hermes data~\cite{Airapetian:2005cb}  for values of $x$ less than 0.2. A postulated model involving hidden-color components of the deuteron is shown to
 complement the effects of pion exchange in reproducing the Hermes data for all measured values of $x$. Predictions are made for an upcoming JLab experiment~\cite{Jlab}. The sum-rule of Close \& Kumano, \eq{cksr}   is shown to be violated for the  three previously published mechanisms that contribute to $b_1$. 
However, the sum rule holds when the mechanism involves valence quarks, such as in the present hidden color model. This means that such contributions (if non-zero) must
yield negative and positive contributions to $b_1$. Finding such an up-down pattern is an  interesting and significant  problem for experimentalists. A clear observation of such a pattern would provide significant evidence for  the existence of hidden-color components of the deuteron.

  \section*{Acknowledgements} This   work has been partially supported by 
U.S. D. O. E.  Grant No. DE-FG02-97ER-41014. I thank M. Alberg for useful discussions and M. Aicher~\cite{Aicher:2010cb} 
 for providing a computer program for $q^\pi$.

 \end{document}